\documentstyle[graphicx,multicol,prl,aps,epsf]{revtex}
\begin{document}
\title{Shot noise in superconducting 
junctions with weak link formed by Anderson impurity}
\author{Yshai Avishai$^{1,2}$,
 Anatoly Golub$^1$ and Andrei D. Zaikin$^{3,4}$ }
\address{$^1$ Department of Physics, Ben-Gurion University of the Negev,
Beer-Sheva, Israel\\
$^2$ NTT Basic Research Laboratory, 3-1 Morinosato Wakamiya, Japan\\
$^3$ Forschungszentrum Karlsruhe, Institut f\"ur Nanotechnologie,
76021 Karlsruhe, Germany\\
$^4$ I.E.Tamm Department of Theoretical Physics, P.N.Lebedev
Physics Institute, 117924 Moscow, Russia}

\maketitle

\begin{abstract}
A theory is developed to study shot noise in superconducting 
($SAS$) and hybrid ($SAN$) junctions with singly occupied
Anderson impurity ($A$) as a weak link. 
The zero-frequency $dc$ component
of the shot noise spectral density 
is calculated at zero temperature
as a  
function of the bias at different 
Coulomb repulsion strengths ($U$), and show a remarkable 
structure resulting from combination of electron-electron 
interaction and Andreev reflections. 
\end{abstract}
\begin{multicols}{2}
\narrowtext

\noindent
{\bf Motivation and main results}: Following the first theoretical work 
on noise in superconducting point junctions
\cite{hlus} the underlying physics
attracted a considerable
theoretical \cite{been,blatter,averin,shumeiko,cuevas,nagaev} 
and experimental \cite{tarasov,jehl,misaki,diel,hoss} attention
(see review article \cite{rev} for more complete list of references).
In most of these works, the quantum point 
contacts have been considered in both the
linear and  nonlinear regimes.
Tunneling junctions in which
the electrodes are separated by a resonant 
double barrier were studied in Refs. \cite{blatter,hlus2}. 
Novel experimental techniques 
now enable the study transport and shot noise in quantum 
dots located between either normal ($N$) or 
superconducting ($S$) electrodes. 
So far, however, the physics of shot noise in superconducting 
junctions with strong electron correlations has not 
been exposed. Investigation of
this fundamental aspect is carried out
in this work. 
The shot noise spectral density in $SAS$ 
and $SAN$ junctions is calculated below for 
the case were the Anderson type impurity 
level ($A$) is singly occupied.
For an $SAN$ junction at small bias 
voltages $V$, doubling of the normal Poisson noise to current 
ratio (Fano-factor) is preserved,
although its dependence on the electron-electron
interaction is quite 
essential. In an $SAS$ junction,  the main process
contributing to the current and shot noise power is
multiple Andreev reflections ($MAR$). 
However, in spite of the fact that for low $V$
the number 
$n\approx 2\Delta/eV$ of $MAR$ is 
large ($\Delta$ is the superconductor gap), the current and shot noise 
density are rather weak. This is due to the low 
effective transparency $\Gamma$ of the junction as a 
consequence of interaction (Coulomb blockade). 
Large-$n$ processes are therefore damped as $\Gamma^{n}$.

\noindent
{\bf Model and effective action}: For convenience, 
the junction is represented 
by two half electrodes on the left
($L$) and the right ($R$) separated by quantum dot.
The dot (located at the origin) is modeled as an Anderson impurity $A$ 
with level position $\epsilon_{0}<0$ 
and Hubbard repulsion $U$ under 
the condition for single occupancy $U>-\epsilon_{0}>0$.
 This can be justified 
by noticing that in recent experiments \cite{sara,kastner}
on semiconductor quantum dots it was 
shown that
tunneling  takes place  through a separate state
with the sign of a Kondo behavior (the tunable Kondo effect). 

The starting point is the tunnel Hamiltonian 
$H=H_{L}+H_{R}+H_{d}+H_{t}$ in which $H_{j} (j=L,R)$ are 
lead Hamiltonians (usually of $BCS$ form) defined in terms of electron 
field operators $\psi_{j \sigma}(x,y,t)$, 
$H_{d}=\epsilon_{0} \sum_{\sigma}c_{\sigma}^{\dagger}
 c_{\sigma} + U n_{\uparrow} n_{\downarrow}$ 
and
$H_{t}={\cal T} \sum_{j\sigma} c^{\dagger}_{\sigma} 
\psi_{j \sigma}({\bf 0},t) + hc$. 
Formally, the whole physics 
is contained in the partition function $Z \equiv \int d[F] exp (iS)$ 
where the path integral Grassman integration is carried out over all 
fermion fields $[F]$ and the action $S$ is obtained by integrating 
the Lagrangian pertaining to the Hamiltonian $H$ along a 
Keldysh contour.  
In a recent work\cite{us} we studied
$I-V$ characteristics in $SAS$ and $SAN$ 
junctions, and developed a formalism to
carry out integration over the fields 
$\psi_{j \sigma}(x,y,t)$, combined with the dynamical 
mean field approximation to account for the quartic 
term $U n_{\uparrow} n_{\downarrow}$.
At the end of this procedure one arrives at an 
effective action $S_{eff}$ which depends on a 
parameter $\gamma_{-}$ whose physical meaning is 
the difference between spin up and spin down 
energies of the dot, (to be determined 
self consistently in terms of the effective action). 
The latter is defined in terms of lead Green functions, 
$g_{j}^{R/A/K}$ where $j=L,R$ and $R/A/K$ denotes 
advanced, retarded and Keldysh respectively. 
Explicitly, dropping the lead index, 
\begin{eqnarray}
&&\hat g=\left ( \begin{array}{cc}
\hat {g^R} & \hat {g^K}\\
 0 & \hat {g^A}
\end{array}\right),
\label{Eq_gmat}
\end{eqnarray}
is a matrix in Keldysh space with
\begin{eqnarray}
&&\hat g^{R/A}(\epsilon )= \frac{(\epsilon 
\pm i0)\tau_z +i|\Delta|\tau_y}
{\sqrt{(\epsilon \pm i0)^2-|\Delta|^2}}, \\
\label{Eq_g0RA}
&&\hat g^K(\epsilon )=
(\hat g^R(\epsilon )-\hat g^A(\epsilon ))
\tanh (\epsilon /2T).
\label{Eq_g0K}
\end{eqnarray}
The set of Pauli matrices $\tau_{x,y,z}$ acts in 
spin (Nambu) space, and the lead dependence enters 
through the corresponding superconducting gaps $\Delta$. 
It is now possible to define the kernel, 
\begin{eqnarray}
&&\hat{G}^{-1}(\epsilon,\epsilon')=\delta
(\epsilon-\epsilon')(\epsilon
 +\gamma_{-}-\tau_z\tilde{\epsilon})+
i\Gamma \hat g_{+}(\epsilon,\epsilon')\tau_z,
\label{Eq_Ginv}
\end{eqnarray} 
(where $\hat {g}_{+} \equiv (\hat {g}_{L}+\hat {g}_{R})/2$) 
as an operator in Nambu $\otimes$ Keldysh product spaces 
(here $\tilde{\epsilon} \equiv \epsilon_{0}+U/2$ 
and $\Gamma \propto {\cal T}^{2}$ is the usual 
transparency parameter). The effective 
action is a functional of the Grassman fields $c, \bar{c}$ 
which are four tuples in this space, 
\begin{eqnarray}
S_{\rm eff}=\int \frac{d\epsilon}{2\pi} \int
d\epsilon'\bar{c} \hat{G}^{-1}(\epsilon,\epsilon')c.
\label{Seff2}
\end{eqnarray}
The self-consistency equation for $\gamma_{-}$ is,
\begin{equation}
\gamma_{-}=\frac{U}{2}\ll\int dt \bar{c}\sigma_{x}c\gg,
\label{Eq_gamma}
\end{equation}
where $\ll O \gg \equiv 
\frac {\int d[c] O exp (iS_{eff})} {\int d[c] exp (iS_{eff})}$ and 
the set of Pauli matrices $\sigma_{x,y,z}$ acts in Keldysh space.

The physical
justification for using 
the mean field approximation 
is related to the fact that the the Kondo resonance 
is suppressed
by the superconducting order parameter. Such 
suppression occurs due to strong attenuation of
the density of states  in an energy region 
of order $\Delta$ near the Fermi energy. 
Thus the number of low energy 
electrons which are able to screen the local impurity spin 
is small \cite{fazio,ambe}. Therefore, we can expect 
that ineffective screening results in a larger domain 
in parameter space 
($\epsilon_0 , U$) for which the single 
occupancy (doublet state) becomes the ground 
state of the whole system. This state is obtained 
from the solution of the mean field equation (\ref{Eq_gamma}).

\noindent
{\bf Explicit expression for shot noise}:
The noise spectrum measures the current 
fluctuations in the junction. It is defined by 
the symmetrized current-current correlation 
function which, 
in terms of current operators in 
Nambu $\otimes$ Keldysh space reads,
\begin{eqnarray}&&
 K(t_{1},t_{2})=\hbar [\hat {T}< I^{(1)}(t_{1})I^{(2)}(t_{2})> \nonumber \\
&&+<\hat {T} I^{(1)}(t_{2})I^{(2)}(t_{1})>-2<I>^2],
\label{Eq_K12}
\end{eqnarray}
where $\hat {T}$ is the time ordering operator and
$<...>$ denotes quantum mechanical thermodynamic averaging 
with respect to the total Hamiltonian $H$.
Starting from the general definition of 
the current operator for tunneling 
through a quantum dot\cite{meir}, in the present case it reads, 
\begin{eqnarray}&&
I^{(1,2)}=\pm \frac{ie}{\hbar}\sum_{k}{ \cal T}
[\bar{c}\frac{1\pm \sigma_{x}}{2}\psi_{Rk}(0)-h.c], 
\label{Eq_I12}
\end{eqnarray}
where $\psi_{Rk}(0)$ is the Fourier transform of 
$\psi_{R}(0,y,t)$ (the so called surface field 
operator) with respect to $y$. 
Substituting the explicit form for the
current operators (\ref{Eq_I12}) into equation (\ref{Eq_K12}) 
we then obtain an expression for $K(t_{1},t_{2})$ 
which involves Grassman integration over 
surface fields and dot electron 
operators. The first integration is Gaussian
and can be done exactly, involving the 
Green function matrix $\hat {g}$.
Integrations over 
the dot fermion fields is completed within the 
dynamic mean field approximation. After 
Fourier transform on $t_{1}-t_{2}$ 
it is possible to express the power spectrum
$K(\omega)$ and $\gamma_{-}$ in terms of the
 Green functions of the entire
system $\hat{G}^{R,A}$ 
(inverse of the kernel defined in Eq.(\ref{Eq_Ginv})).
In the present work, attention is focused on the
zero-frequency $dc$ component of the 
shot noise $K=K(0)$ at 
zero temperature. 
The first novel result of our study then
consists of a workable expression for the
noise power density function  
$ K=K_{1}+K_{2}$ supported by a 
self-consistent equation for 
the energy occupancy parameter $\gamma_{-}$, that is, 
\begin{eqnarray}
K_{1}= \frac{i e^2 \Gamma}{2\hbar} 
\int \frac{d \epsilon}{2 \pi}Tr \{ \tau_{z}[
(\hat{g}^{R}_{R}-\hat{g}^{A}_{R})
(\hat{G}^{R}-\hat{G}^{A})-\hat{g}^{K}_{R}F] \},
\label{Eq_K1}
\end{eqnarray}
\begin{eqnarray}&&
K_{2}=\frac{e^2 \Gamma^{2}}{8\hbar } 
\int \frac{d \epsilon}{2\pi}
Tr \{\frac{1}{2}(F\tilde{g})^{2}
-\tau_{z}\tilde{g}\tau_{z}
\hat{G}^{A}\tilde{g}\hat{G}^{R}- 
\nonumber \\
&& 2\tilde{g}\hat{G}^{R}
\hat{g}^{K}_{R}F+(\hat{G}^{R}\hat{g}^{K}_{R})^{2}-
\tilde{g}^{R}\hat{G}^{R}
\tilde{g}^{R}\hat{G}^{R}+h.c] \},
\label{Eq_K2}
\end{eqnarray}
\begin{eqnarray}
&& \gamma_{-}=-i\frac{U}{2}
\int \frac{d \epsilon}{2 \pi}Tr F,
\label{Eq_gammaminus}
\end{eqnarray}
where,
\begin{eqnarray}&&
F=\frac{ -i \Gamma}{2 }\hat{G}^{R}
(\hat{g}^{K}_{R}+\hat{g}^{K}_{L})
\tau_{z}\hat{G}^{A},
\label{Eq_F}
\end{eqnarray}
\begin{eqnarray}
&&\tilde{g}
=\tau_{z}(\hat{g}^{R}_{R}
+(\hat{g}^{R}_{R})^{+})\tau_{z},
\label{Eq_tildeg}
\end{eqnarray}
\begin{eqnarray} 
&&\tilde{g}^{R,A}=\tau_{z}\hat{g}^{R}_{R,A}
\tau_{z}-\hat{g}^{R}_{R,A}.
\label{Eq_gRA}
\end{eqnarray}
We arrive now at our main goal, namely, using  expressions 
(\ref{Eq_K1}-\ref{Eq_gRA}) to analyze 
the noise in $SAN$ and $SAS$ junctions.
 Before  starting this analysis we
note that Eqs. (\ref{Eq_K1},\ref{Eq_K2},\ref{Eq_gammaminus}) 
can be easily 
generalized for the case with different left 
and right transparency parameters: $\Gamma_{R}, \Gamma_{L}$. 
To do this we write  Eqs. (\ref{Eq_K1},\ref{Eq_K2})
 in a symmetric form. Moreover,
Eq. (\ref{Eq_F}) is to be replaced by 
$F=\frac{ -i}{2 }\hat{G}^{R}
(\Gamma_{R}\hat{g}^{K}_{R}+\Gamma_{L}\hat{g}^{K}_{L})
\tau_{z}\hat{G}^{A}$.

\noindent
{\bf Shot noise in $SAN$ junctions}: In this case 
only one Andreev reflection takes place and
$K_{2}$ can be written 
explicitly as an integral 
over the whole energy domain (including sub-gap), while
for $K_{1}$, integration is effected above $\Delta$ only.
From that part of $K_{1}+K_{2}$ 
related to integration
 with $|\epsilon|>\Delta$ we 
easily obtain the shot noise for $NAN$
 junction,
(tunneling between normal metal electrodes
through an Anderson-impurity center).
It assumes
a standard form \cite{hlus}, although 
it depends on the interaction through the 
parameter $\gamma_{-}$, that is,  
\begin{eqnarray}
&& K=\frac{e^3|V|}{\pi\hbar}[T_{-}
(1-T_{-})+T_{+}(1-T_{+})],
\label{Eq_KSAN}
\end{eqnarray}
where $T_{\pm}=\Gamma^2/((\tilde 
{\epsilon} \pm \gamma_{-})^2+\Gamma^2 )$. 
In a perfect resonance condition
($\tilde \epsilon=0$ and $U=0$) we get $K=0$ 
as expected for
a pure point junction \cite {hlus,rev}.
When $U \ne 0$
we have to solve the
self-consistent equation (\ref{Eq_gammaminus}) to find 
$\gamma_{-}$ and substitute the
 solution for a given voltage into equations 
(\ref{Eq_K1},\ref{Eq_K2}). 

Another limiting situation for which a
simple analytical expression exists
is that of small 
bias $eV<\Delta$. For completely 
transparent point junctions
the current-noise spectral density 
vanishes \cite{hlus}. For non-resonant
transport there is a finite shot noise in 
the sub-gap region which, in the limit 
of low effective transparency,
yields the Fano factor  to 
be equal to $2$. The principal contribution 
to the noise density in the regime
 $eV\ll \Delta$ comes from $K_{2}$. Therefore, integrating 
over the sub-gap energies in equations 
(\ref{Eq_K1},\ref{Eq_K2}, \ref{Eq_gammaminus}) results in,
\begin{eqnarray}
&& K=\frac{2e^3|V|}{\pi\hbar}
\frac{\Gamma^4[\Gamma^2 \tilde 
\epsilon^2+(\gamma_{-}^2-\tilde 
\epsilon^2)^2]}
{[\Gamma^2\gamma_{-}^2+(\tilde 
\epsilon^2+0.5\Gamma^2 -\gamma_{-}^2)^2]^2}.
\label{Eq_K}
\end{eqnarray}
This formula, together with the
self-consistency equation (\ref{Eq_gammaminus}) 
yields the shot noise 
density  for $T=0$ and $eV \ll \Delta $. 
For $\Gamma \ll \tilde{\epsilon},\gamma_{-}$ it
leads to twice the normal Poisson noise-current ratio. 
Numerical
 calculation using the exact expressions 
(\ref{Eq_K1},\ref{Eq_K2}) for specific values of $U$ 
indicates that the   
deviation from the approximate relation (\ref{Eq_K})) starts 
already at $eV\sim\Delta/2$. 
Therefore, even in the sub-gap 
regime, the non-linear $V$-dependence 
of the shot noise is important and 
exact expressions for the spectral density 
should be employed. 
The shot noise of an $SAN$ junction is
displayed in Fig.1
(hereafter, all the energies
are expressed  in units of 
$\Delta$ ).  
\begin{figure}[htb]
\includegraphics[width=0.35\textwidth,keepaspectratio]{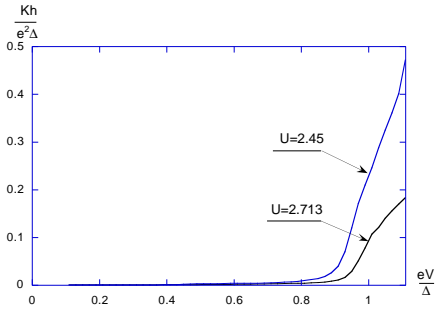}
\caption
{The shot noise $K$ versus the bias $V$
for an $SAN$ junction at sub-gap voltages.
The parameters are $\epsilon_0 =-1.5$,
$\Gamma=0.6$, $U=2.45$  and 2.713.}
\label{figure1}
\end{figure}
For the
parameters employed in the figure, the noise in the sub-gap regime 
is rather small, starting to grow near 
the superconducting gap. 
It is strongly dependent on the Coulomb interaction
which, as a general rule, suppresses it
in comparison with its value appropriate
for simple tunneling through a non-interacting 
impurity. It is interesting to note that the role of the repulsive 
Hubbard constant $U$ in an $SAN$ junction (in the 
regime of single occupancy) is similar to that 
of an exchange term in a contact between a superconductor 
and a ferromagnetic metal. Therefore, 
our results might be relevant for this junction as well. 
Note also  that for clean point junctions, 
 there is a saturation value for the shot 
noise spectrum density at large voltages \cite{hlus}. 
This value is readily reproduced from our 
formulas (\ref{Eq_K1},\ref{Eq_K2}) if we consider the limit 
$\Gamma \gg \Delta$ at resonance ($\tilde{\epsilon}=0$)
for bias voltage $eV>\Delta$. The only  
contribution to the correlation function 
comes from energy integration above the gap 
in equations (\ref{Eq_K1},\ref{Eq_K2}), yielding 
the saturated value,
\begin{eqnarray}
K_{ex}=\frac{4e^2}{\pi\hbar}lim_{eV \to
\infty} \{\int_{\Delta}^{eV} d\epsilon 
\frac{N(\epsilon)-1}{(N(\epsilon)+1)^2} \}
\;=\; \frac{4e^2\Delta}{15\pi\hbar},
\label{Eq_Kex}
\end{eqnarray}
where $N(\epsilon)=\frac {|\epsilon|}{\sqrt{\epsilon^{2}-\Delta^{2}}}$.

\noindent
{\bf Shot noise in $SAS$ junctions}:
The current-noise spectral density of an
$SAS$ junction is calculated 
numerically from equations (\ref{Eq_K1},\ref{Eq_K2}).
For a constant bias $V$ it is useful 
to discretize the energy integration\cite{us}. 
The energy domain in equations 
(\ref{Eq_K1},\ref{Eq_K2},\ref{Eq_gammaminus})
is divided into slices 
of width $2eV$ and
integration is performed on an interval
$[0<E<2eV]$. The Green functions become
matrices with indices representing 
the different energy slices. 
Explicitly,
\begin{eqnarray}
 (m| \hat G^{-1}(\epsilon)|n) &=& 
\delta_{m,n}\left[\epsilon_{m}+
\gamma_{-}-
\tau_{z}\tilde{\epsilon}+
\frac{ i\Gamma}{2}\hat g_{R}
(\epsilon_{m})\tau_{z}\right]
+\nonumber\\
& &\frac { i\Gamma}{2} (m|\hat g_{L}
(\epsilon )|n)\tau_{z},
\label{Eq_Minv1}
\end{eqnarray}
\begin {eqnarray}
(m|\hat g_{L}(\epsilon)|n)&=&
(\hat g^{11}_{L}(\epsilon_{m}^{-})P_{+}+
g^{22}_{L}(\epsilon_{m}^{+})P_-)\delta_{m,n}
+\nonumber\\
& &g^{12}_{L}(\epsilon_{m}^{-}
)\tau_{+}\delta_{n,m-1}+
g^{21}_{L}(\epsilon_{m}^{+} )\tau_{-}\delta_{n,m+1},
\label{Eq_GEE}
\end{eqnarray}
where $\epsilon_{m}=\epsilon+2meV$, $\epsilon_{m}^{\pm}=\epsilon_{m}\pm eV$,  
the superscripts denote the matrix 
elements in Nambu space and
$P_{\pm}=(1\pm \tau_z)/2$. We dropped the constant phase difference
$\phi_{0}=arg(\Delta_{L}-\Delta_{R})$ which
does not play a special role here.

We are mainly interested
in the voltage dependence at the sub-gap 
region where $MAR$ are 
important and the way how this noise-voltage
characteristic is influenced by Coulomb interaction.
The result is displayed in Fig.2. 
\begin{figure}[htb]
\includegraphics[width=0.35\textwidth,keepaspectratio]{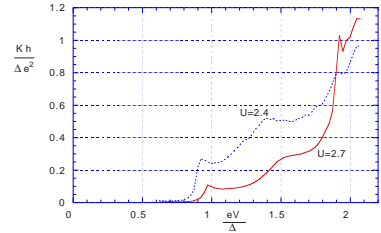}
\caption
{The shot noise $K$ versus bias $V$
for an $SAS$ junction at sub-gap voltages.
The parameters are $\epsilon_0 =-1.5$,
$\Gamma=0.6$, $U=2.4$  and 2.7.}
\label{figure2}
\end{figure}
As can be deduced,
 there is a strong effect of the Coulomb 
repulsion
on the noise spectrum $K$ in the sub-gap region. 
For voltages not 
too close to $2 \Delta$ the noise is smaller 
for the higher value of $U$. This is similar 
to what happens with the current itself, as can be judged
by a glance at the $I-V$ curve (Fig. 3). 
\begin{figure}[htb]
\includegraphics[width=0.35\textwidth,keepaspectratio]{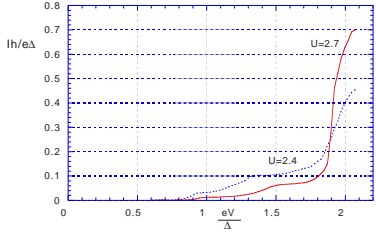}
\caption{$I-V$
curves for an $SAS$ junction at sub-gap
voltages. Parameters are the same as in Fig.2}
\label{figure3}
\end{figure}
And yet, the Fano 
factor $K/2eI$  is nearly 
independent on interaction (see Fig.4). This finding
stresses the importance of $MAR$ process,  analogous to what
has been argued in junctions with low transparency \cite{cuevas}. 
\begin{figure}[htb]
\includegraphics[width=0.35\textwidth,keepaspectratio]{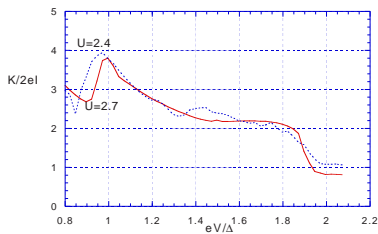} 
\caption
{The Fano factor $K/2eI$ versus bias
$V$ for an $SAS$ junction at sub-gap
voltage. The parameters are as in Fig.2}
\label{figure4}
\end{figure}
As in the case of SAN junction there is a 
saturated value for the spectral noise 
density power also for SAS junctions. 
In the limit $\Gamma \gg \Delta$ this saturated 
value is twice of that of Eq.(\ref{Eq_Kex}).

In conclusion, based on a
theory  developed for the study of 
tunneling in $SAS$ and $SAN$ junctions,
\cite{us}, the shot noise in
such systems is calculated.
Special attention is devoted to 
analyzing
the implications of the Coulomb
 repulsion between 
electrons in the dot on the tunneling 
process in general and the noise spectrum in particular. 
The theoretical treatment uses a 
combination 
of Keldysh non-equilibrium Green 
function 
and path integral formalism, and the interaction 
is handled within the
dynamical mean field approximation. 
General expressions for the 
current noise spectral density correlation 
function are then derived.

The main results of the present research 
can be summarized as follows: 1). In 
$SAN$ and $SAS$ junctions, the Coulomb interaction 
results in a nonzero value for the occupancy 
energy ($\gamma_{-}$) and thus always
acts as a factor which lowers the
transparency. Therefore in $SAN$
 junctions the shot noise  appears already at 
$eV<\Delta$ (even if $ \tilde{\epsilon}=0 $).
The doubled value of the Fano factor emerges
in the limit $\Gamma \ll \tilde{\epsilon},\gamma_{-}$.
 2) In $SAS$ junctions,
the shot 
noise correlation function
in the sub-gap regime ($eV<2\Delta $) 
is suppressed at higher values of $U$.
The Fano coefficient is nearly 
interaction independent, and unambiguously 
manifests the importance of $MAR$.

\noindent
This research is supported by DIP 
German Israel Cooperation project {\bf Quantum
electronics in low dimensions} 
by the Israeli Science Foundation 
grants {\bf Center of Excellence} and {\bf Molecular bridges} and 
by the US-Israel BSF grant {\bf Dynamical instabilities 
in quantum dots}.

\end{multicols}

\begin{thebibliography}{99} 
\bibitem{hlus} V. A. Khlus, Zh. Eskp. 
Teor. Fiz. {\bf 93}, 2179 (1987) 
[Sov. Phys. JETP {\bf 66}, 1243 (1987)].
\bibitem{been} M. J. M. de Jong and C. 
W. J. Beenakker, Phys. Rev. B 
{\bf 49}, 16070 (1994).
\bibitem{blatter} A. L. Fauche're, G. 
B. Lesovik, and G. Blatter, Phys. Rev. B 
{\bf 58}, 11177 (1998).
\bibitem{averin} D. Averin and H. T. 
Imam , Phys. Rev. Lett. {\bf 76}, 3814 (1996).
\bibitem{shumeiko} J. P. Hessling, V. S. 
Shumeiko, Yu. M. Galperin, and G. Wendin 
Europhys. Lett. {\bf 34}, 49 (1996).
\bibitem{cuevas} J. C. Cuevas {\it et al.} 
Phys. Rev. Lett. {\bf 82}, 4086 (1999).
\bibitem{nagaev} K. E. Nagaev and M. 
Buttiker cond-mat/0007121.
\bibitem{tarasov} A. N. Vystavkin and M. A. 
Tarasov, Pis'ma Zh. Tekh. Phys {\bf 9}, 869 
(1983) [Sov. Tech. Phys. Lett. {\bf 9}, 373 (1983)].
\bibitem{jehl}  X. Jehl, {\it et al.}, 
Phys. Rev. Lett. {\bf 83}, 1660 (1999).
\bibitem{misaki} Y. Misaki {\it et al.}, 
Jpn. J. Appl. Phys. Pt.1 {\bf 35}, 1190 (1996).
\bibitem{diel} P. Dieleman {\it et al.} 
Phys. Rev. Lett. {\bf 79}, 3486 (1997).
\bibitem{hoss} T. Hoss {\it et al.} 
cond-mat/9901129.
\bibitem{rev} Ya. Blanter and M. 
Buttiker, cond-mat/9910158.
\bibitem{hlus2} S. V. Naidenov 
andV. A. Khlus, Fiz.Iizk.Temp. {\bf 21}, 594 (1995) 
\bibitem{us} Y. Avishai, A. Golub, and A.D. Zaikin, cond-mat/0007098.
\bibitem{fazio} R. Fazio and R. Raimondi , 
Phys. Rev. Lett. {\bf 80}, 2913 (1998); {\bf 82}, 4950 (E)(1999).
\bibitem{ambe} A. A. Clerk, V. Ambegoakar, and S. Hershfield , 
Phys. Rev. B {\bf 61}, 3555 (2000).
\bibitem{meir} Y. Meir and N. S. Wingreen , 
Phys. Rev. Lett. {\bf 68}, 2512 (1992).
\bibitem{sara} S. M. Cronenwett, T. H. 
Oosterkamp, L. P. Kouwenhovev, Science  
{\bf 281}, 540 (1998).
\bibitem{kastner} D. Goldhaber, J. 
Gores, M. A. Kastner, H. Shtrikman, 
D. Mahalu, ana U. Meirav, Phys. Rev. 
Lett. {\bf 81}, 5225 (1998).
\end{thebibliography}
\end{document}